\documentclass[prd, twoside, 12pt,tightenlines,singlecolumn,nofootinbib, floatfix, showpacs,aps]{revtex4-1}

\usepackage{amsmath,amsfonts,amssymb}
\usepackage{mathrsfs}
\usepackage{graphicx}
\usepackage[english]{babel} 
\usepackage{verbatim}
\usepackage[colorinlistoftodos]{todonotes}
\usepackage[colorlinks=true, allcolors=blue]{hyperref}

\newcommand{\ANLHEP}{HEP Division, Argonne National Laboratory, Lemont, IL 60439, USA}
\newcommand{\APC}{Laboratoire Astroparticule et Cosmologie (APC), CNRS/IN2P3, Universit\'e Paris Diderot, 10, rue Alice Domon et L�onie Duquet, 75205 Paris Cedex 13, France}
\newcommand{\UCLA}{University of California at Los Angeles, Los Angeles,  CA 90095}

\newcommand{\BenGurion}{Department of Physics, Ben-Gurion University, Be'er Sheva 84105, Israel}
\newcommand{\BNL}{Brookhaven National Laboratory, Upton, NY 11973}
\newcommand{\Brown}{Brown University, Providence, RI 02912}

\newcommand{\Cincinnati}{University of Cincinnati, Cincinnati, OH 45221}

\newcommand{\damtp}{DAMTP, Centre for Mathematical Sciences, Wilberforce Road, Cambridge, UK, CB3 0WA}

\newcommand{\UChicago}{University of Chicago, Chicago, IL 60637}

\newcommand{\IBS}{Institute for Basic Science (IBS), Daejeon 34051, Korea}

\newcommand{\ICCD}{Institute for Computational Cosmology, Department of Physics, Durham University, South Road, Durham, DH1 3LE, UK}

\newcommand{\IFUNAM}{IFUNAM - Instituto de F\'{i}sica, Universidad Nacional Aut\'onoma de M\'exico, 04510 CDMX, M\'exico}
\newcommand{\IHEP}{Institute of High Energy Physics, Austrian Academy of Sciences, 1050 Vienna, Austria}

\newcommand{\INAFOAS}{INAF - Osservatorio di Astrofisica e Scienza dello Spazio di Bologna, via Piero Gobetti 93/3, I-40129 Bologna, Italy}

\newcommand{\ITFA}{Institute for Theoretical Physics, University of Amsterdam, Science Park 904, 1098 XH Amsterdam, The Netherlands}

\newcommand{\KASSI}{Korea Astronomy and Space Science Institute, Daejeon 34055, Korea}
\newcommand{\kavli}{Kavli Institute for Cosmology, Cambridge, UK, CB3 0HA}
\newcommand{\KICP}{Kavli Institute for Cosmological Physics, Chicago, IL 60637}
\newcommand{\KIPAC}{Kavli Institute for Particle Astrophysics and Cosmology, Stanford 94305}

\newcommand{\Kobe}{Kobe University, 657-8501 Kobe, Japan}

\newcommand{\LBL}{Lawrence Berkeley National Laboratory, Berkeley, CA 94720}
\newcommand{\Leiden}{Lorentz Institute, Leiden University, Niels Bohrweg 2,Leiden, NL 2333 CA, The Netherlands}

\newcommand{\LLNL}{Lawrence Livermore National Laboratory, Livermore, CA, 94550}

\newcommand{\MIT}{Massachusetts Institute of Technology, Cambridge, MA 02139}

\newcommand{\OSU}{The Ohio State University, Columbus, OH 43212}
\newcommand{\OU}{Department of Physics and Astronomy, Ohio University, Clippinger Labs, Athens, OH 45701, USA}

\newcommand{\SimonFraser}{Department of Physics, Simon Fraser University, Burnaby, British Columbia, Canada V5A 1S6}

\newcommand{\INFNFE}{Istituto Nazionale di Fisica Nucleare, Sezione di Ferrara, 40122, Italy }

\newcommand{\SLAC}{SLAC National Accelerator Laboratory, Menlo Park, CA 94025}

\newcommand{\StonyBrook}{Stony Brook University, Stony Brook, NY 11794}

\newcommand{\UCBP}{Department of Physics, University of California Berkeley, Berkeley, CA 94720, USA}

\newcommand{\UCD}{University of California at Davis, Davis, CA 95616}
\newcommand{\UCI}{University of California, Irvine, CA 92697}

\newcommand{\UGTO}{Instituto de F\'{i}sica de la Universidad de Guanajuato, A. P. 150, C. P. 37150, Le\'{o}n, Guanajuato, Mexico}

\newcommand{\Umich}{University of Michigan, Ann Arbor, MI 48109}
\newcommand{\UMN}{University of Minnesota, Minneapolis, MN 55455}

\newcommand{\UNIPD}{Dipartimento di Fisica e Astronomia ``G. Galilei'',Universit\`a degli Studi di Padova, via Marzolo 8, I-35131, Padova, Italy}

\newcommand{\UoM}{Jodrell Bank Center for Astrophysics, School of Physics and Astronomy, University of Manchester, Oxford Road, Manchester, M13 9PL, UK}

\newcommand{\UWMadison}{Department of Physics, University of Wisconsin - Madison, Madison, WI 53706}

\newcommand{\VSI}{Van Swinderen Institute for Particle Physics and Gravity, University of Groningen, Nijenborgh 4, 9747~AG~Groningen, The~Netherlands}

\begin{document}
\begin{center}
{\bf\LARGE Astro2020 Science White Paper}\\
{\bf\Large Gravitational probes of ultra-light axions}
\end{center}
{\bf Primary thematic area:} Cosmology and Fundamental Physics\\ \\
\noindent {\bf Principal author:} Daniel Grin\\
\noindent \textbf{e-mail:} dgrin@haverford.edu\\
\noindent \textbf{Institution:} Haverford College\\
\noindent \textbf{Phone number:} (610) 896-2908\\ \\
\noindent\textbf{Authors:} \\ \noindent
Mustafa~A.~Amin$^{1}$, {Vera~Gluscevic$^{2,3}$, Daniel~Grin$^{4}$, Ren\'{e}e~Hl\v{o}zek$^{5,6}$,  David~J.~E.~Marsh$^{7}$, Vivian~Poulin$^{8,9}$, Chanda~Prescod-Weinstein$^{10}$, Tristan~L.~Smith$^{11}$}\\ \\
\noindent
\textbf{Abstract:}\\
The axion is a hypothetical, well-motivated dark-matter particle whose existence would explain the lack of charge-parity violation in the strong interaction. In addition to this original motivation, an `axiverse' of ultra-light axions (ULAs) with masses $10^{-33}\,{\rm eV}\lesssim m_{\rm a}\lesssim 10^{-10}\,{\rm eV}$ also emerges from string theory. Depending on the mass, such a ULA contributes to the dark-matter density, or alternatively, behaves like dark energy. At these masses, ULAs' classical wave-like properties are astronomically manifested, potentially mitigating observational tensions within the $\Lambda$CDM paradigm on local-group scales. ULAs also provide  signatures on small scales such as suppression of structure, interference patterns and solitons  to distinguish them from heavier dark matter candidates. Through their gravitational imprint, ULAs in the presently allowed parameter space furnish a host of observational tests to target in the next decade, altering standard predictions for microwave background anisotropies, galaxy clustering, Lyman-$\alpha$ absorption by neutral hydrogen along quasar sightlines, pulsar timing, and the black-hole mass spectrum.\newpage
\noindent\textbf{Endorsers (affiliation list follows after references):}
\\ \noindent
Zeeshan~Ahmed$^{12}$, Eric~Armengaud$^{13}$, Robert Armstrong$^{14}$, Carlo Baccigalupi$^{15}$, \\Marco~Baldi$^{16,17,18}$, Nilanjan Banik$^{19,20}$, Rennan Barkana$^{21}$, Darcy Barron$^{22}$, \\Daniel Baumann$^{20,23}$, Keith Bechtol$^{24}$
% ...
\normalsize
\section{Axions: Motivation \& Background}
The nature of dark matter (DM) and dark energy (DE) that dominate the energy density of our universe remains a mystery. Axions are hypothetical particles proposed in the 1970s \cite{Peccei:1977hh,Weinberg:1977ma,Wilczek:1977pj}, and could constitute significant fractions of the DM and DE. They could resolve challenges to the $\Lambda$ cold dark matter (CDM) cosmology on small scales and impact cosmological observables \cite{Hu:2000ke}, like the CMB. Axions are characterized by a mass $m_{\rm a}$ measured in ${\rm eV}$, and `decay constant' $f_{\rm a}$, which determines the importance of axion self-interactions.  Beyond their original motivation to solve the charge-parity symmetry problem of Quantum Chromodynamics (QCD), axions also arise in many ``beyond the standard model'' (BSM) theories of particle physics, such as string theory \cite{Svrcek:2006yi,Acharya:2010zx,Cicoli:2012sz}. There could be many ($\sim 200$) axion species \cite{Demirtas:2018akl}, and evidence for even one could hint at a larger ``axiverse''  \cite{Arvanitaki:2009fg,Marsh:2015xka,Hui:2016ltb,Stott:2017hvl,Visinelli:2018utg}.

For DM and DE to have the right abundances, the parameter ranges \cite{Marsh:2015xka}
\begin{equation}
    10^{-33}\text{ eV}\leq m_{\rm a}\leq 10^{-2}\text{ eV};\quad 10^{7}\text{ GeV}\leq f_{\rm a} \leq 10^{18}\text{ GeV}\, .
\end{equation} are of particular interest. In these ranges, axions are produced non-thermally, thus evading hot DM bounds.\footnote{There may also be a thermal sub-population, tested by measurements of $N_{\rm eff}$ \cite{Baumann:2016wac}.} 
Early on, when the Hubble parameter $H\lesssim mc^{2}/\hbar$, axions have equation-of-state parameter $w\simeq -1$ and their density scales as $\rho\approx {\rm constant}$ \cite{Frieman:1995pm,Marsh:2010wq}. Afterwards, the axion field begins to rapidly oscillate, and on average, $w\simeq 0$, with $\rho\propto (1+z)^{3}$ \cite{Abbott:1982af,PhysRevD.28.1243}, where $z$ is the cosmological redshift. Depending on when this transition occurs, axions can contribute to either the DM or DE of the universe. An axion with $m_{\rm a}\lesssim10^{-27}\text{ eV}$ will behave as early DE, while one with $m_{\rm a}\lesssim 10^{-33}~{\rm eV}$ behaves as standard DE and can drive cosmic acceleration today. Heavier axions behave as DM. 

A variety of experimental and observational techniques could detect axions, such as radio-frequency resonance techniques \cite{Rosenberg:2010zz,Lewis:2017gno}, nuclear magnetic resonance methods \cite{Graham:2013gfa,Budker:2013hfa}, telescope searches \cite{Raffelt:2006cw}, and others \cite{Vogel:2013bta,TheMADMAXWorkingGroup:2016hpc,Du:2018uak,Marsh:2018dlj,Irastorza:2018dyq}. These rely on axion couplings to standard model (SM) particles, which scale as $f_{\rm a}^{-1}$ \cite{Sikivie:1983ip,Marsh:2015xka}. Direct detection relies on a large cosmic density $\Omega_{\rm a}$ and knowledge of the local DM density. Laboratory searches probe the ranges $10^{-17}\text{ eV}\lesssim m_{\rm a}\lesssim 10^{-2}\text{ eV}$ and $10^{9}\text{ GeV}\lesssim f_{\rm a} \lesssim 10^{16}~\text{GeV}$.  

Axions would enhance stellar cooling \cite{Raffelt:2006cw,Cadamuro:2012rm,Friedland:2012hj,Perna:2012wn,Giannotti:2015kwo,Giannotti:2017hny}, affecting stellar populations and asteroseismology \cite{Raffelt:2006cw,Isern:2008nt,Cadamuro:2012rm}. Axions can also be detected through their gravitational signatures if their Compton wavelength $\lambda_C$ is astronomically relevant. If $m_{\rm a}\simeq 10^{-10}\text{ eV}$, $\lambda_C$ is comparable to the Schwarzschild radius of a stellar mass black hole (BH). Axions could then spin down BHs \cite{Arvanitaki:2009fg,Kitajima:2018zco}. If $m_{\rm a}\simeq 10^{-33}\text{ eV}$, $\lambda_C$ is comparable to the cosmic horizon, and axions could cause present-day cosmic acceleration. These scales bracket the \emph{gravitational ultra-light axion (ULA) window}, where it is possible to search for axions using only their gravitational interactions, with no reliance on their (highly suppressed) couplings to the SM.  Gravitational effects are entirely complementary to other signatures. We refer to all such particles as ULAs, noting the existence of other terms, like axion-like particles (ALPs) and weakly interacting slim particles (WISPs) \cite{Arias:2012az}. We focus on real fields like the axion, but we note that complex scalar fields with $m_{\rm a}\simeq 10^{-22}~{\rm eV}$ are also a DM candidate \cite{Li:2016mmc}. They behave as axions at early and late times, but redshift as $\rho\propto (1+z)^{6}$ and then $(1+z)^{4}$ at intermediate times, with distinct CMB and gravitational-wave signatures \cite{Li:2016mmc}.

Cosmological observations are uniquely able to probe the ULA relic density, $\Omega_{\rm a}$ (which is set by $m_{\rm a}$, $f_{\rm a}$, and dimensionless initial field value $\theta_{i}$). If $f_{\rm a} \sim 10^{16}~{\rm GeV}$, the energy scale of Grand Unified Theories or GUTs, and $\theta_i\approx 1$, ULA densities could be cosmologically relevant.
For comparison's sake, we note that CMB and large-scale structure observations have established that the DE density is $\Omega_\Lambda=0.685\pm 0.007$, the DM density is $\Omega_c=0.264\pm 0.007$, and relic neutrinos must have $\Omega_\nu\leq 0.005$ \cite{Aghanim:2018eyx}. For masses in the range $10^{-33}\text{ eV}\lesssim m_{\rm a}\lesssim 10^{-16}\text{ eV}$ the relic density is in the range $10^{-4}\lesssim \Omega_{\rm a}\lesssim 1$, as shown in Fig.~\ref{fig1}. Larger ULA densities correspond to even larger values of $f_{\rm a}$, approaching the Planck scale, $2.4\times 10^{18}\text{ GeV}$. If some of the dark sector is composed of ULAs, then, the cosmological observables test low $m_{\rm a}$ and large $f_{a}$ values, well beyond the reach of other techniques. The GUT prediction for $\Omega_{\rm a}$ motivates exploration of the entire ULA mass window. We treat $\Omega_{\rm a}$ and $m_{\rm a}$ as free parameters, keeping an open mind to a wide range of theoretical scenarios.

\vspace{-0.2 in}
\section{Axions and challenges to $\Lambda$CDM}
\vspace{-0.1 in}
Let us not only ask what astrophysics can do for axions, but also what can axions do for astrophysics. ULAs could explain late-time cosmic acceleration, or compose some of the DM \cite{Matos:1992qx,Matos:1999et,Hu:2000ke,Suarez:2011yf,Hwang:2009js,Park:2012ru,Magana:2012xe}. They could cause an early DE epoch \cite{Poulin:2018cxd}, thus mitigating tensions between CMB and local measurements of the Hubble constant $H_{0}$. ULAs recover the successes of pure $\Lambda$CDM on large scales but alleviate challenges to $\Lambda$CDM on small scales \cite{Hu:2000ke,Bullock:2017xww}. Some dwarf galaxy DM halo profiles exhibit cores in their central density profiles, in contrast with naive $\Lambda$CDM predictions \cite{Bullock:2017xww}. If ULAs exist and compose a significant fraction of DM, their macroscopic de Broglie wavelengths would cause galactic halo density profiles to have cores at $\sim 0.7~(m_{\rm a}/10^{-22}~{\rm eV})^{-1/2}~{\rm kpc}$ scales, possibly explaining observations of cores in Milky-Way dwarf Spheroidal galaxies \cite{Lee:1995at,Hu:2000ke,Arbey:2003sj,Marsh:2013ywa,Marsh:2015wka,Gonzales-Morales:2016mkl,Bernal:2017oih,Deng:2018jjz,Broadhurst:2019fsl}. In particular, a dynamical analysis of Fornax and Sculptor yields `hints' of a ULA solution to CDM small-scale problems with $0.3\times 10^{-22}~{\rm eV}\lesssim m_{\rm a}\lesssim 10^{-22}~{\rm eV}$ \cite{Marsh:2015wka}.

By suppressing the growth of cosmic structure on the smallest scales, ULAs in this mass range could address challenges to the $\Lambda$CDM paradigm, such as the paucity of Milky-Way satellite galaxies compared to expectations, by suppressing the number of low-mass subhalos around Milky-Way scale halos. ULAs in this window would also address the ``too big to fail problem", the surprising depression of satellite galaxy masses in halos relative to $\Lambda$CDM expectations \cite{Marsh:2013ywa,Marsh:2015xka,Marsh:2016vgj,Hui:2016ltb}. The region of interest is shown as ``ULA hints" in Fig.~1, where we also show current constraints and possibilities for new probes. The properties of the ultra-faint dwarf galaxy Eridanus II exclude most of the ULA mass window $0.8\times 10^{-21}~{\rm eV}\lesssim m_{\rm a}\lesssim 10^{-19}~{\rm eV}$ \cite{Marsh:2018zyw}. There are other  tests of ULA DM using galactic dynamics \cite{Robles:2015gwa,Martinez-Medina:2015jra,Emami:2018rxq,Li:2018kyk}.\footnote{There are also baryonic explanations for these puzzles, and they will soon be critically tested.} 

Axions may behave as a coherent macroscopic low-momentum state wave, referred to as Bose-Einstein condensate (BEC) dark matter \cite{Sikivie:2009qn,Erken:2011xj,Saikawa:2012uk,Guth:2014hsa,2015arXiv150105913B,Banik:2015xda}. Other ultra-light particles can display similar behavior, although the mass scale will determine the astrophysical phenomenology \cite{Rindler-Daller:2013zxa,Li:2013nal}.  If axions thermalize gravitationally, there could be galaxy-scale BECs \cite{Sikivie:2009qn}, yielding caustic rings that are testable using galaxy rotation curves \cite{Natarajan:2005ut,Duffy:2008dk,Rindler-Daller:2013zxa,Dumas:2015wba,Chakrabarty:2018gdg}. The self-interaction of these particles is meaningful in determining the scale of stability of the BEC \cite{Guth:2014hsa}. Axions have \textit{attractive} self-interactions at leading order and thus fail one requirement for stable large-scale Bose condensation, the presence of \textit{repulsive} self-interactions \cite{Guth:2014hsa,Chavanis:2016dab,Berges:2017ldx,Levkov:2018kau}. This could lead to much-smaller coherence lengths, producing solitons or ``axion stars" and ``axteroids"\cite{Chavanis:2011cz,Guth:2014hsa,Eby:2014fya,Eby:2015hyx,Eby:2016cnq,Visinelli:2017ooc,Chavanis:2017loo,Amin:2019ums} (with mass $\sim 10^{-11} M_{\odot}$ for QCD axions). For ULAs, solitons are more massive, $\sim 10^{6}\to 10^{9}~M_{\odot}$, and could constitute dwarf galaxy density cores \cite{Schwabe:2016rze,Hui:2016ltb}, with implications for halo structure and detection prospects. We explore the opportunities for astrophysical ULA detection in the next decade, starting with the CMB.

\vspace{-0.2 in}
\section{Axions and the CMB}
\begin{figure}
    \centering
    \includegraphics[width=.85\textwidth]{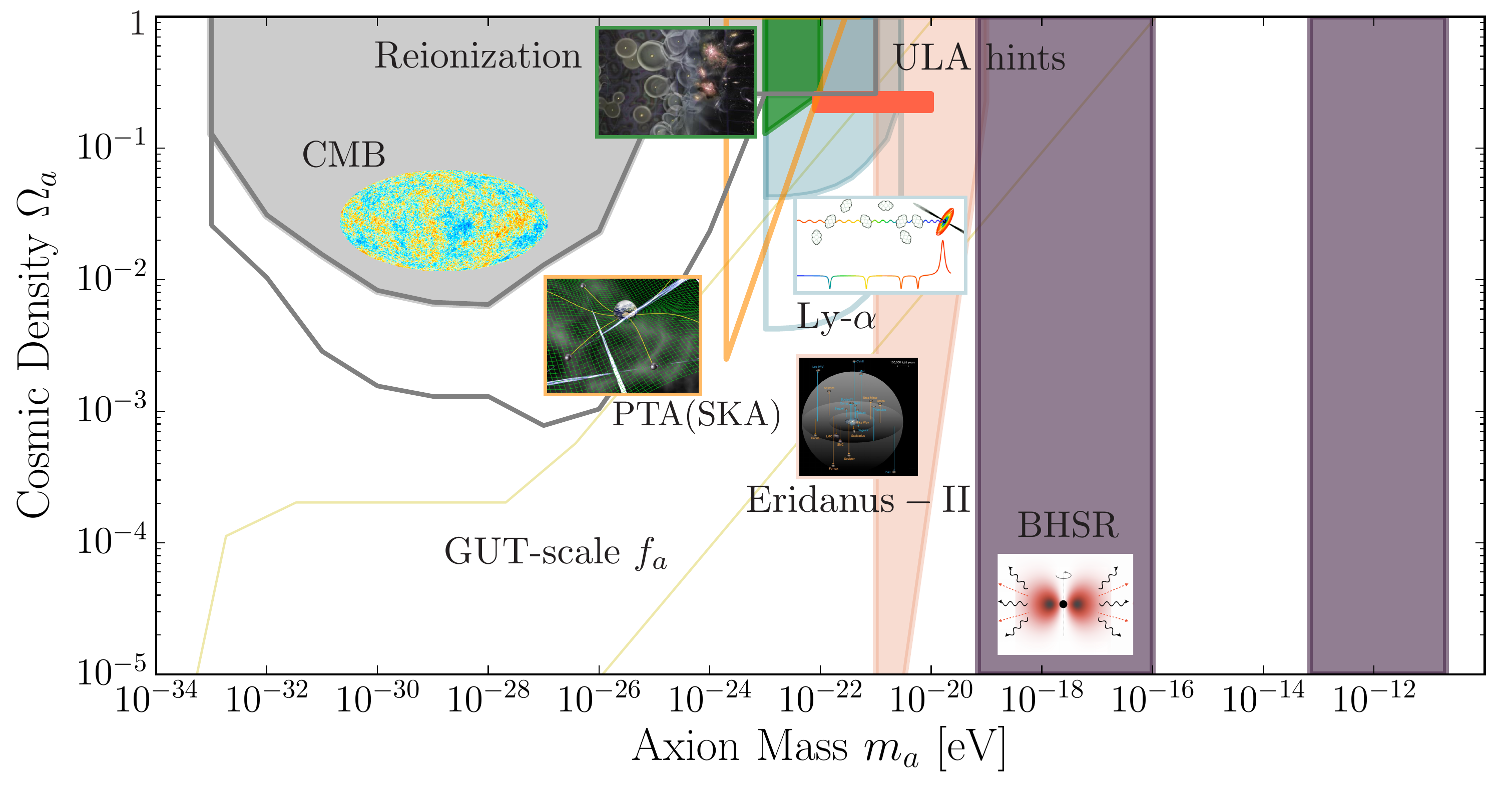}
    \caption{{The cosmic window on ultralight axions, showing the reach of various astronomical probes. Shaded regions are currently excluded. Lines below them indicate the sensitivity of future experiments/surveys. ULA hints refers to the ULA mass scale suggested by MW-scale challenges. Local group schematic, \textit{Credit: J. T. A. de Jong, Leiden University.} Planck CMB map, \textit{Credit ESA, http://www.esa.int}.   Black-hole super-radiance schematic \cite{Arvanitaki:2010sy} used with permission from American Physical Society, License RNP/19/MAR/012767. PTA Schematic \textit{Image Credit: David Champion.} Lyman Alpha Schematic, \textit{Image Credit: Ned Wright.} Reionization schematic used with permission from artist, J.~F.~Podevin, originally used in Ref. \cite{2006SciAm.295e..46L}. \label{fig1}}}
\end{figure}

In the DE-like mass window, ULAs roll slowly down their potential as a dark-energy component,
shifting CMB acoustic peaks to smaller angular scales (higher $\ell$), and increasing the largest scale anisotropies due to gravitational potential-well decay for $0\lesssim z\lesssim 3300$ \cite{Frieman:1995pm,Coble:1996te,Marsh:2010wq,Marsh:2013taa,Hlozek:2014lca,Emami:2016mrt}. In the DM-like mass window, the imprint of ULAs on the Hubble expansion and perturbation growth alters peak heights measured in the temperature and E-mode polarization power spectra \cite{Amendola:2005ad,Park:2012ru,Marsh:2013taa,Hlozek:2014lca,Urena-Lopez:2015gur}. ULAs manifest wave-like properties on cosmological scales, suppressing density fluctuations the ULA comoving ``Jeans scale"
$\lambda\lesssim \lambda_{\rm J}\equiv 0.1~{\rm Mpc}~(m_{\rm a}/10^{-22}\,{\rm eV})^{-1/2}(1+z)^{1/4}$ \cite{Khlopov:1985jw,Hu:2000ke,Hwang:2009js,Chavanis:2011uv,Park:2012ru,Hlozek:2014lca,Suarez:2015fga,Marsh:2015xka,Urena-Lopez:2015gur,Hlozek:2016lzm,Cedeno:2017sou,Hlozek:2017zzf}, affecting comoving wavenumbers $k>2\pi/\lambda_{\rm J}$. This affects the strength and scale-dependence of gravitational lensing of the CMB \cite{Seljak:1998aq,Zaldarriaga:1998te,Okamoto:2003zw,Hanson:2009kr}.

In the window $10^{-32}\,{\rm eV}\lesssim m_{\rm a}\lesssim 10^{-26}\,{\rm eV}$, the imprint of these effects and \textit{Planck} satellite data impose the constraint of $\Omega_{\rm a}\lesssim 10^{-2}$, as shown in Fig.~\ref{fig1} \cite{Hlozek:2017zzf,Poulin:2018dzj}. Fig.~\ref{fig1} includes the effect of CMB lensing, which improves sensitivity by a factor of $\sim 3$ compared with the unlensed CMB or galaxy clustering. 
For $m_{\rm a}\lesssim 10^{-32}\,{\rm eV}$ (or $m_{\rm a}\gtrsim 10^{-25}\,{\rm eV})$, ULA effects on CMB/galaxy clustering signatures are indistinguishable from a cosmological constant (or DM) with current data, lifting these constraints \cite{Hlozek:2014lca}. 

In the next and coming decades, very sensitive experiments like the Simons Observatory (SO) \cite{Ade:2018sbj}, CMB Stage-4 (CMB-S4) \cite{Abazajian:2016yjj}, and PICO \cite{Hanany:2019lle} (e.g., map noise levels of $6~\mu{\rm K}$-${\rm arcmin}$, $1~{\mu}{\rm K}$-${\rm arcmin}$, and $0.6~{\mu}{\rm K}$-${\rm arcmin}$, respectively) will achieve nearly cosmic-variance limited measurements of CMB primary anisotropies, and reduce lensing reconstruction noise by a factor of $\sim 20$.  
CMB-S4 should improve ULA sensitivity to $\Omega_{\rm a}\sim 10^{-3}$ at the most constrained $m_{\rm a}$ values and probe the range $m_{\rm a}\sim 10^{-23}\,{\rm eV}$ \cite{Hlozek:2016lzm}. CMB lensing will drive future tests of BSM scenarios \cite{Hlozek:2016lzm}. Additionally, electromagnetic couplings of ULAs will produce additional signatures from CMB anisotropies, including CMB spectral distortions \cite{Lee:2014rpa,Sigl:2018fba,Mukherjee:2018oeb,Mukherjee:2018zzg,Fedderke:2019ajk}.

Unique among DM candidates, axions with $f_{\rm a}\gtrsim H_I$ imply the presence of a low-mass field during the inflationary era, exciting DM (isocurvature) fluctuations that are statistically independent from baryons, neutrinos, and radiation \cite{Axenides:1983hj,Turner:1983sj,Lyth:1991ub,Fox:2004kb,Hertzberg:2008wr,Komatsu:2008hk,Langlois:2010xc}. This would cause a phase shift and low-$\ell$ amplitude change in the CMB peaks \cite{Baumann:2015rya}. The size of this ULA imprint would be controlled by $H_{I}$. $H_{I}$ sets the amplitude of the inflationary gravitational-wave (GW) background, which is detectable through CMB B-mode polarization, parameterized by the tensor-to-scalar ratio $r$ \cite{Seljak:1996gy,Kamionkowski:1996ks}. In the mass window $10^{-25}~{\rm eV}\lesssim m_{\rm a}\lesssim 10^{-24}~{\rm eV}$, current data allow a $\sim 10\%$ contribution of ULAs to the DM along with $\sim 1\%$ contributions of isocurvature and tensors to the CMB power spectrum \cite{Hlozek:2016lzm}.

Detecting primordial B-modes is a science driver for ground-based CMB observatories and future space missions (e.g. Spider \cite{Gualtieri:2017zcz}, BICEP2-3/Keck Array \cite{Grayson:2016smb,Ade:2018gkx}, CLASS \cite{2014SPIE.9153E..1IE}, Simons Array \cite{Westbrook:2018vod}, SO \cite{Galitzki:2018avl}, CMB-S4 \cite{Abazajian:2016yjj}, LiteBIRD \cite{2018JLTP..193.1048S}), and PICO \cite{Hanany:2019lle}). These efforts could probe values $r\sim 10^{-4}\to 10^{-3}$. Cosmological probes of $\Omega_{\rm a}$ and  the inflationary energy scale \cite{Visinelli:2009zm,Marsh:2013taa,Marsh:2014qoa,Visinelli:2014twa,Visinelli:2017imh} are thus complementary. For $\Omega_{\rm a}$ saturating current bounds, the combined isocurvature and lensing sensitivity of CMB-S4 would probe values $H_{\rm I}\gtrsim 10^{13.3}~{\rm GeV}$ \cite{Abazajian:2016yjj,Hlozek:2017zzf}. A detection of primordial B-modes at S4 sensitivity levels would test ULA contributions at the level of $\Omega_{\rm a}\simeq 0.01$ \cite{Abazajian:2016yjj,Hlozek:2017zzf}. Foreground cleaning techniques may even allow measurements of the lensing power spectrum to multipoles of $L\sim 40,000$ \cite{Nguyen:2017zqu}. The CMB could then distinguish between pure CDM and ULAs, in the preferred $\sim 10^{-22}\,{\rm eV}$ window hinted at in the Milky Way \cite{Marsh:2013ywa,Marsh:2016vgj,Bullock:2017xww}. This window will be tested by other structure formation probes.
\vspace{-0.25 in}
\section{Optical \& infrared surveys}\vspace{-.1 in}Another probe of axions is the clustering power spectrum $P_{g}(k)$ of galaxies, which has already imposed the limit $\Omega_{\rm a}\sim 3\times 10^{-2}$ at the most constrained $m_{\rm a}$ values, comparable to the unlensed CMB alone \cite{Blake:2010xz,Hlozek:2014lca}. The Large Synoptic Survey telescope (LSST) will produce surveys that improve error bars in $P_{g}(k)$ as well as weak lensing observables and thus sensitivity to $\Omega_{\rm a}$ by an order of magnitude \cite{Abell:2009aa}. Additionally, future space missions like Euclid and WFIRST will improve sensitivity to the matter power-spectrum $P_{\delta}(k)$ by an order of magnitude \cite{Amendola:2016saw,Thompson:2013ufa,Spergel:2013uha,Dore:2018smn}. Lyman-$\alpha$ (Ly$\alpha$) absorption of quasar emission depends on the optical depth of neutral hydrogen, and is sensitive to DM densities at early times \cite{Armengaud:2017nkf,Desjacques:2017fmf,Kobayashi:2017jcf,Irsic:2017yje,Nori:2018pka,Leong:2018opi}. Depending on modeling details, the data impose a limit of $m_{\rm a}\gtrsim 1\to 40 \times 10^{-22}\,{\rm eV}$, if ULAs compose all of the DM.

Next-generation spectroscopic surveys like the Dark Energy Spectroscopic Instrument (DESI), Euclid and WFIRST will improve sensitivity to the matter power-spectrum $P_{\delta}(k)$ from Ly$\alpha$ measurements by an order of magnitude \cite{Aghamousa:2016zmz}. The small-scale power in ULA models relative to $\Lambda$CDM models scales as $(\lambda/\lambda_{J})^{8}\propto m_{\rm a}^{4}$, and so these measurements could push the ULA mass limit as high as $m_{\rm a}\gtrsim 4\to 160 \times 10^{-22}\,{\rm eV}$. There are also prospects for Milky-Way scale effects. LSST's \textit{complete} census of ultra-faint dwarf galaxies in the Milky Way will test extensions to standard $\Lambda$CDM, including ULAs \cite{Abell:2009aa}. An ultra-faint dwarf census can put an upper limit on $M_{\rm half}$, the mass scale below which r.m.s density fluctuations fall to half their $\Lambda$CDM values \cite{Drlica-Wagner:2019xan}. LSST could test $M_{\rm half}$ values of $10^{6}M_{\odot}$, probing values $m_{\rm a}\gtrsim 10^{-19}~{\rm eV}$ \cite{Drlica-Wagner:2019xan}. 

The reionization of the universe occurred around redshift $z\sim 6$ and is tied to structure formation at early times and small scales. CMB constraints to the Thomson scattering optical depth from reionization and measurements of the galaxy luminosity function depend on the low mass galaxy population, and would be impacted by a significant fraction of ULA DM \cite{Bozek:2014uqa}. Current observations allow the range $m_{\rm a}\geq 10^{-22}~{\rm eV}$. Future James Webb Space Telescope (JWST) observations will constrain the UV galaxy luminosity function to rest-frame magnitudes of $M_{\rm UV}=-16$, beyond the $M_{\rm UV}=-17$ cutoff of $10^{-22}~{\rm eV}$ ULA DM, testing the ULA explanation for challenges to $\Lambda$CDM. Significant sensitivity will come from the non-linear regime \cite{Markovic:2010te}, and so comparisons between N-body \cite{Zhang:2016uiy}, fluid \cite{Nori:2018hud}, Schr\"{o}dinger \cite{Schive:2014dra,Schive:2014hza,Mocz:2017wlg,Du:2018zrg,Veltmaat:2018dfz} and Klein-Gordon approaches are essential, as is consideration of self-interactions \cite{Amin:2011hu,Cedeno:2017sou,Desjacques:2017fmf,Poulin:2018dzj,Leong:2018opi}.
\vspace{-.25in}
\section{Black holes and gravitational wave astronomy}\vspace{-.1 in}
In the coming decade, pulsar timing arrays could detect a stochastic background of GWs down to $\sim 10^{-9}~{\rm Hz}$ or individual $\sim 10^{9}M_{\odot}$ 
supermassive black hole (BH) binaries. If DM has a significant ULA component, the Milky Way gravitational potential will oscillate [on time scales $\delta t\simeq 20 ~{\rm ns}~(m_{\rm a}/10^{-23}~{\rm eV})^{-3}(\Omega_{\rm a}/0.3)]$,  within reach of pulsar timing efforts like NANOGrav and the Parkes Pulsar Timing Array (if $m_{\rm a}\lesssim 10^{-23}~{\rm eV}$) \cite{Porayko:2018sfa}, and distinct from GW signals. The Square Kilometer Array (SKA) will improve sensitivity to $\Omega_{\rm a}$ by an order of magnitude \cite{Porayko:2018sfa}. Reaching values $m_{\rm a}\gtrsim 10^{-22}~{\rm eV}$ requires hundreds of new millisecond pulsars \cite{Porayko:2018sfa}, though the detailed density profiles of ULAs \cite{DeMartino:2017qsa} could improve sensitivity. Collisions and mergers of ultra-compact solitons could generate gravitational waves in addition to standard sources (BHs and neutron stars) \cite{Palenzuela:2017kcg,Helfer:2018vtq,Clough:2018exo}.

Another powerful phenomenon is BH Superradiance (BHSR) \cite{Arvanitaki:2009fg,Arvanitaki:2010sy,Marsh:2015xka,Stott:2018opm,Kitajima:2018zco}. If ULAs exist, they would form bound states near spinning BHs. An instability would lead to the spin down of BHs with Kerr radii comparable to their ULA Compton wavelengths, while transitions between bound states could lead to GW emission, enhanced in binary systems \cite{Baumann:2018vus}. The existence of stellar and supermassive BH populations rules out ULAs with $6\times 10^{-13}~{\rm eV}\lesssim m_{\rm a}\lesssim 2\times 10^{-11}~{\rm eV}$ and $10^{-18}~{\rm eV} \lesssim m_{\rm a} \lesssim 10^{-16}~{\rm eV}$ \cite{Arvanitaki:2010sy,Stott:2018opm}. Future measurements of the BH mass function by Advanced LIGO could push the lower limit of the stellar mass ULA window down to $1\times 10^{-13}~{\rm eV}$. In the Advanced LIGO band, detections are possible in the $m_{\rm a}\sim 10^{-10}~{\rm eV}$ range, while the LISA band is sensitive to values $m_{\rm a}\sim 10^{-17}~{\rm eV}$ \cite{Kitajima:2018zco}.

\vspace{-0.2 in}
\section{Conclusions \& Recommendations}
\vspace{-0.1 in}

Ultra-light axions could be an important component of the dark sector, as late-time dark energy, early dark energy, or dark matter, depending on the axion mass. This merits attention in the next decade, especially if searches for weakly interacting massive particles (WIMPs) yield null results \cite{Bertone:2018xtm}. We make the following recommendations:\begin{itemize}\itemsep0em \item Support proposed CMB \cite{Galitzki:2018avl,Abazajian:2016yjj,Hanany:2019lle} and large-scale structure surveys \cite{Abell:2009aa,Amendola:2016saw,Thompson:2013ufa,Spergel:2013uha,Dore:2018smn}.
\item Support independent investigations in theory, simulations and data analysis, as well as comparisons of diverse methods via interaction between groups worldwide. 
\item Support development of novel probes, from ultra small-scale CMB experiments \cite{Nguyen:2017zqu} to X-ray studies of neutron star cooling \cite{2019arXiv190303035R} and very large samples of pulsars \cite{Porayko:2018sfa}, so that the window for all ultra-light axion dark matter $10^{-23}~{\rm eV}\lesssim m_{\rm a}\lesssim 10^{-19}~{\rm eV}$ is decisively probed via diverse methods, on the small scales where discovery potential is greatest.
\end{itemize}

\newpage
\bibliographystyle{apsrev4-1} 

\newpage
\begin{center}\textbf{Affiliations}\end{center}
\noindent{\scriptsize
$^{1}$Department of Physics \& Astronomy, Rice University, Houston, Texas 77005, USA\\
$^2$Department of Physics, University of Florida, Gainesville, FL 32611\\
$^{3}$Department of Physics, Princeton University, Princeton, NJ 08544\\
$^4$Haverford College, 370 Lancaster Ave, Haverford PA, 19041, USA\\
$^{5}$Department of Astronomy and Astrophysics, University of Toronto, M5S 3H4\\
$^6$Dunlap Institute for Astronomy and Astrophysics, University of Toronto, 50 St.~George Street, ON M5S 3H4, Canada\\
$^7$Institut f\"{u}r Astrophysik, Georg-August Universit\"{a}t, Friedrich-Hund-Platz 1, D-37077 G\"{o}ttingen, Germany\\
$^8$Laboratoire Univers \& Particules de Montpellier (LUPM), CNRS \& Universit\'e de Montpellier (UMR-5299),Place Eug\`ene Bataillon, F-34095 Montpellier Cedex 05, France\\
$^{9}$Dept. of Physics and Astronomy, The Johns Hopkins University, Baltimore, MD, USA 21218\\
$^{10}$Department of Physics and Astronomy, 9 Library Way, University of New Hampshire, Durham, NH 03824\\
$^{11}$Department of Physics and Astronomy, Swarthmore College, 500 College Ave, Swarthmore, PA, 19081, USA\\
$^{12}$\SLAC\\
$^{13}$IRFU, CEA, Universit\'{e} Paris-Saclay, F-91191 Gif-sur-Yvette, France\\
$^{14}$\LLNL\\
$^{15}$SISSA, Astrophysics Sector, via Bonomea 265, 34136, Trieste, Italy\\
$^{16}$Dipartimento di Fisica e Astronomia, Alma Mater Studiorum Universit\'{a} di Bologna, via Gobetti 93/2, I-40129 Bologna, Italy\\
$^{17}$\INAFOAS\\
$^{18}$INFN - Sezione di Bologna, viale Berti Pichat 6/2, I-40127 Bologna, Italy\\
$^{19}$\Leiden \\
$^{20}$GRAPPA Institute, Institute for Theoretical Physics Amsterdam and Delta Institute for Theoretical Physics, University of Amsterdam, Netherlands\\
$^{21}$Raymond and Beverly Sackler School of Physics and Astronomy, Tel Aviv University, Tel Aviv 69978, Israel\\
$^{22}$Department of Physics and Astronomy, University of New Mexico, 1919 Lomas Blvd NE, Albuquerque, New Mexico 87131, USA\\
$^{23}$\ITFA\\
$^{24}$\UWMadison\\
$^{25}$\Cincinnati \\
$^{26}$\ANLHEP\\
$^{27}$\KICP \\
$^{28}$Canadian Institute for Theoretical Astrophysics (CITA), University of Toronto, ON, Canada\\
$^{29}$\LBL \\ 
$^{30}$Department of Theoretical Physics, University of The Basque Country UPV/EHU, E-48080 Bilbao, Spain\\
$^{31}$Donostia International Physics Center (DIPC), 20018 Donostia, The Basque Country\\
$^{32}$IKERBASQUE, Basque Foundation for Science, E-48013 Bilbao, Spain\\
$^{33}$\UChicago \\
$^{34}$\UCBP \\
$^{35}$\OU\\
$^{36}$Department of Physics, Harvard University, Cambridge, MA, 02138, USA\\
$^{37}$\UCI \\
$^{38}$Stanford Institute for Theoretical Physics, Physics Department, Stanford University, Stanford, CA 94306\\
$^{39}$Jet Propulsion Laboratory, California Institute of Technology, Pasadena, CA, USA\\
$^{40}$The Observatories of the Carnegie Institution for Science, 813 Santa Barbara Street, Pasadena, CA 91101, USA\\
$^{41}$Institute for Theory and Computation, Harvard University, 60 Garden Street, Cambridge, MA 02138, USA\\
$^{42}$Astrophysics, University of Oxford, DWB, Keble Road, Oxford OX1 3RH, UK\\
$^{43}$Department of Physics, University of California, San Diego, La Jolla, CA 92093, USA\\
$^{44}$Department of Physics, Kenyon College, 201 N College Rd, Gambier, OH 43022, USA\\
$^{45}$\UGTO\\
$^{46}$The Oskar Klein Centre for Cosmoparticle Physics, Stockholm University, Roslagstullsbacken 21A, SE-106 91 Stockholm, Sweden\\
$^{47}$\UMN\\
$^{48}$Institute of Cosmology, Department of Physics and Astronomy Tufts University, Medford, MA 02155, USA\\
$^{49}$\ICCD\\
$^{50}$School of Natural Sciences, Institute for Advanced Study, 1 Einstein Drive, Princeton, NJ 08540, USA\\
$^{51}$Center for Computational Astrophysics, Flatiron Institute,162 5th Avenue, 10010, New York, NY, USA\\
$^{52}$\OSU\\
$^{53}$Center for Theoretical Physics, Department of Physics, Columbia University, New York, NY 10027\\
$^{54}$\Umich\\
$^{55}$University of Washington, Department of Astronomy, 3910 15th Ave NE, WA 98195-1580 Seattle \\
$^{56}$\IBS\\
$^{57}$\KASSI\\
$^{58}$\UCD\\
$^{59}$\Brown \\
$^{60}$\BenGurion\\
$^{61}$\INFNFE\\
$^{62}$Tsinghua Center for Astrophysics and Department of Physics, Tsinghua University, Beijing 100084, China\\
$^{63}$Department of Physics and Astronomy, University of Pennsylvania, 209 South 33rd Street, Philadelphia, PA 19104, USA\\
$^{64}$\UNIPD\\
$^{65}$\IFUNAM\\
$^{66}$Departamento de F\'{i}sica, Centro de Investigaci\'{o}n y de Estudios Avanzados del IPN, A.P. 14-740, 07000 CDMX, M\'{e}xico\\
$^{67}$\MIT\\
$^{68}$Department of Physics, Cornell University, Ithaca, NY 14853, USA\\
$^{69}$\kavli \\
$^{70}$\damtp \\
$^{71}$\VSI \\
$^{72}$Department of Physics, 3215 Daniel Ave, Southern Methodist University Dallas, Texas 75205\\
$^{73}$International Centre for Theoretical Physics, Trieste, Italy\\
$^{74}$Institut d'Astrophysique de Paris, UMR 7095, CNRS, 98 bis boulevard Arago, F-75014 Paris, France\\
$^{75}$\BNL\\
$^{76}$INRIM, Strada delle Cacce 91, I-10135 Torino, Italy\\
$^{77}$\SimonFraser\\
$^{78}$\IHEP\\
$^{79}$\KIPAC\\
$^{80}$Theoretical Physics Department, CERN, 1 Esplanade des Particules, CH-1211 Geneva 23, Switzerland\\
$^{81}$Max-Planck-Institut f\"{u}r Physik (Werner-Heisenberg-Institut), F\"{o}hringer Ring 6, 80805 M\"{u}nchen, Germany\\
$^{82}$Berkeley Center for Theoretical Physics, Department of Physics, University of California, Berkeley, CA 94720, USA\\
$^{83}$Departamento de F\'{i}sica T\'{e}orica, Universidad de Zaragoza, Pedro Cerbuna 12, E-50009, Zaragoza, Espa\~{n}a\\
$^{84}$Institut f\"{u}r Astrophysik, Universit\"{a}tssternwarte Wien, University of Vienna, A-1180 Vienna, Austria\\
$^{85}$Institute of Astrophysics, National Taiwan University, 10617 Taipei, Taiwan\\
$^{86}$\StonyBrook\\
$^{87}$Department of Astronomy, University of Texas, Austin, TX 78712-1083, USA\\
$^{88}$\Kobe\\
$^{89}$Department of Astrophysical Sciences, Princeton University, Princeton, NJ 08540, USA\\
\scriptsize
$^{90}$Theoretical Astrophysics Group, Fermi National Accelerator Laboratory, Batavia, IL 60510, USA\\
$^{91}$\APC \\
$^{92}$\UoM \\
$^{93}$NASA Goddard Space Flight Center, Greenbelt, MD 20771, USA\\
$^{94}$Department of Physics, Syracuse University, Syracuse, NY 13244, USA\\
$^{95}$\UCLA \\}

\end{document}